\documentclass{article}
\usepackage{amssymb}
\begin{document}
\title{The Liouville theorem as a problem of common eigenfunctions}

\author{G.F.\ Torres del Castillo \\ Departamento de F\'isica Matem\'atica, Instituto de Ciencias \\
Universidad Aut\'onoma de Puebla, 72570 Puebla, Pue., M\'exico}

\maketitle

\begin{abstract}
It is shown that, by appropriately defining the eigenfunctions of a function defined on the extended phase space, the Liouville theorem on solutions of the Hamilton--Jacobi equation can be formulated as the problem of finding common eigenfunctions of $n$ constants of motion in involution, where $n$ is the number of degrees of freedom of the system.
\end{abstract}

\section{Introduction}
In the framework of the Hamiltonian formulation of classical mechanics, the Liouville theorem asserts that, for a mechanical system with $n$ degrees of freedom, if we have $n$ constants of motion in involution, $F_{1}, F_{2}, \ldots, F_{n}$ (that is, $\{ F_{i}, F_{j} \} = 0$ for $i, j = 1, 2, \ldots, n$, where $\{ \; , \; \}$ is the Poisson bracket), then a complete solution of the Hamilton--Jacobi (HJ) equation can be found by quadrature \cite{Wi,GV,BB,FM,DB,Li}.

More precisely, if $F_{1}(q_{i}, p_{i}, t), \ldots, F_{n}(q_{i}, p_{i}, t)$ are $n$ constants of motion in involution, that is,
\begin{equation}
\frac{\partial F_{i}}{\partial t} + \{ F_{i}, H \} = 0, \qquad i = 1, 2, \dots, n \label{cm}
\end{equation}
and
\begin{equation}
\{ F_{i}, F_{j} \} = 0, \qquad i, j = 1, 2, \dots, n, \label{inv}
\end{equation}
where $\{ \;, \; \}$ denotes the Poisson bracket (with the convention $\{ q_{i}, p_{j} \} = \delta_{ij}$), then, assuming that
\begin{equation}
\det \left( \frac{\partial F_{i}}{\partial p_{j}} \right) \not= 0, \label{reg}
\end{equation}
so that, locally at least, we can express the $p_{i}$ as functions of $q_{j}, F_{j}$, and $t$, the differential form
\begin{equation}
p_{i}(q_{j}, F_{j}, t) \, {\rm d} q_{i} - H \big( q_{i}, p_{i}(q_{j}, F_{j}, t), t \big) \, {\rm d} t \label{df}
\end{equation}
is the differential of some function, $S(q_{i}, t)$, which is a complete solution of the HJ equation (here and in what follows, there is summation over repeated indices).

The aim of this paper is to show that the Liouville theorem can be formulated in another form, closer to the standard formalism of quantum mechanics. Specifically, we shall show that if $S(q_{i}, t)$ is a common eigenfunction of the functions $F_{1}, F_{2}, \ldots, F_{n}$ (a concept to be defined below) then, by adding to $S$ an appropriate function of $t$ only, one obtains a complete solution of the HJ equation.

In Section 2 we present the definition of the eigenfunctions of a function $f(q_{i}, p_{i}, t)$ and we show that two functions, $f(q_{i}, p_{i}, t)$ and $g(q_{i}, p_{i}, t)$, have common eigenfunctions if and only if their Poisson bracket vanishes. Then, we prove that, if conditions (\ref{cm})--(\ref{reg}) hold, a common eigenfunction of $F_{1}, F_{2}, \ldots, F_{n}$ is, up to an additive function of $t$ only, a complete solution of the HJ  equation. In Section 3 we give some illustrative examples, emphasizing the fact that we can make use of constants of motion that depend explicitly on the time.

The statement of the Liouville theorem presented here allows us to see that the Liouville theorem is analogous to one of the methods employed to solve the Schr\"odinger equation, where we look for the common eigenfunctions of a complete set of mutually commuting operators that also commute with the Hamiltonian (e.g., for a spherically symmetric Hamiltonian we consider the common eigenfunctions of $L^{2}$, $L_{z}$ and $H$).

\section{Eigenfunctions of a function and complete solutions of the Hamilton--Jacobi equation}
We start by giving the definition of the eigenfunctions of a real-valued function $f(q_{i}, p_{i}, t)$: We shall say that $S(q_{i}, t)$ is an eigenfunction of $f(q_{i}, p_{i}, t)$, with eigenvalue $\lambda$, if $S$ is a solution of the first-order partial differential equation
\begin{equation}
f(q_{i}, \frac{\partial S}{\partial q_{i}}, t) = \lambda. \label{eig}
\end{equation}
(It may be noticed that if $f$ is a time-independent Hamiltonian, then (\ref{eig}) is the corresponding time-independent HJ equation.) We note that if $S(q_{i}, t)$ is an eigenfunction of $f(q_{i}, p_{i}, t)$ with eigenvalue $\lambda$, then so it is $S(q_{i}, t) + \phi(t)$, for any function $\phi(t)$ of $t$ only, and that the solutions of (\ref{eig}) will depend parametrically on $\lambda$. Of course, in order for (\ref{eig}) to be a differential equation, $f$ must depend on one of the $p_{i}$, at least.

For instance, according to this definition, the eigenfunctions of the function
\begin{equation}
F(q, p, t) = m \omega q \sin \omega t + p \cos \omega t, \label{cons}
\end{equation}
where $m$ and $\omega$ are constants, are the solutions of the differential equation
\[
m \omega q \sin \omega t + \frac{\partial S}{\partial q} \cos \omega t = \lambda,
\]
which can be readily integrated giving
\begin{equation}
S = \lambda q \sec \omega t - \frac{m \omega}{2} q^{2} \tan \omega t + \phi(t),
\end{equation}
where $\phi(t)$ is an arbitrary function of $t$ only. Note that $\lambda$ may be a function of $t$.

If $S(q_{i}, t)$ is a common eigenfunction of $f(q_{i}, p_{i}, t)$ and $g(q_{i}, p_{i}, t)$, with eigenvalues $\lambda$ and $\mu$, respectively, that is, $S$ satisfies (\ref{eig}) and
\[
g(q_{i}, \frac{\partial S}{\partial q_{i}}, t) = \mu,
\]
then, differentiating with respect to $q_{i}$, making use of the chain rule, we obtain
\[
\frac{\partial f}{\partial q_{i}} + \frac{\partial f}{\partial p_{j}} \frac{\partial^{2} S}{\partial q_{j} \partial q_{i}} = 0 \quad {\rm and} \quad \frac{\partial g}{\partial q_{i}} + \frac{\partial g}{\partial p_{j}} \frac{\partial^{2} S}{\partial q_{j} \partial q_{i}} = 0,
\]
hence,
\begin{eqnarray*}
\{ f, g \} & = & \frac{\partial f}{\partial q_{i}} \frac{\partial g}{\partial p_{i}} - \frac{\partial g}{\partial q_{i}} \frac{\partial f}{\partial p_{i}} = - \frac{\partial f}{\partial p_{j}} \frac{\partial^{2} S}{\partial q_{j} \partial q_{i}} \frac{\partial g}{\partial p_{i}} + \frac{\partial g}{\partial p_{j}} \frac{\partial^{2} S}{\partial q_{j} \partial q_{i}} \frac{\partial f}{\partial p_{i}} \\
& = & \left( \frac{\partial^{2} S}{\partial q_{j} \partial q_{i}} - \frac{\partial^{2} S}{\partial q_{i} \partial q_{j}} \right) \frac{\partial f}{\partial p_{i}} \frac{\partial g}{\partial p_{j}} = 0.
\end{eqnarray*}
Thus, if $f(q_{i}, p_{i}, t)$ and $g(q_{i}, p_{i}, t)$ possess common eigenfunctions, then $\{ f, g \} = 0$.

In order to see that the converse is also true, we now assume that $\{ f, g \} = 0$. If $f$ and $g$ are functionally independent, then there exists, locally at least, a set of canonical coordinates, $Q_{i}, P_{i}$, such that, $P_{1} = f$ and $P_{2} = g$. Then, the eigenvalue equations for $f$ and $g$ are $\partial S/\partial Q_{1} = \lambda$ and  $\partial S/\partial Q_{2} = \mu$, which have the simultaneous solutions $S = \lambda Q_{1} + \mu Q_{2} + \phi(Q_{3}, \ldots, Q_{n}, t)$, where $\phi$ is an arbitrary function of $n - 1$ variables, thus showing that $f$ and $g$ have common eigenfunctions. (Note that this does not mean that every eigenfunction of $f$ is an eigenfunction of $g$ (cf.\ \cite{Sn}, Sec.\ 2.9).) The expression for $S$ in terms of the original coordinates, $(q_{i}, t)$, is not given by the simple substitution of the $Q_{i}$ as functions of $(q_{i}, p_{i}, t)$ \cite{CT}; what is relevant here is the existence of common eigenfunctions for $f$ and $g$.

In the case where $f$ and $g$ are functionally dependent, the eigenvalue equations for $f$ and $g$ are equivalent to each other and, trivially, possess common solutions.

\subsection{Alternative formulation of the Liouville theorem}
We now assume that $F_{1}, \ldots, F_{n}$ are $n$ functions satisfying (\ref{cm})--(\ref{reg}), and we consider a common eigenfunction $S(q_{i}, t)$ of $F_{1}, \ldots, F_{n}$, with eigenvalues $\lambda_{1}, \ldots, \lambda_{n}$, respectively, then, assuming that the eigenvalues are constant, differentiating with respect to $t$ both sides of the equation
\begin{equation}
F_{i}(q_{j}, \frac{\partial S}{\partial q_{j}}, t) = \lambda_{i}, \label{eigf}
\end{equation}
making use of the chain rule, the Hamilton equations and (\ref{cm}), we have
\begin{eqnarray*}
0 & = & \frac{\partial F_{i}}{\partial q_{j}} \dot{q_{j}} + \frac{\partial F_{i}}{\partial p_{j}} \frac{{\rm d}}{{\rm d} t} \frac{\partial S}{\partial q_{j}} + \frac{\partial F_{i}}{\partial t} \\
& = & \frac{\partial F_{i}}{\partial q_{j}} \frac{\partial H}{\partial p_{j}} + \frac{\partial F_{i}}{\partial p_{j}} \left( \frac{\partial^{2} S}{\partial t \partial q_{j}} + \frac{\partial^{2} S}{\partial q_{k} \partial q_{j}} \dot{q}_{k} \right) - \frac{\partial F_{i}}{\partial q_{j}} \frac{\partial H}{\partial p_{j}} + \frac{\partial F_{i}}{\partial p_{j}} \frac{\partial H}{\partial q_{j}} \\
& = & \frac{\partial F_{i}}{\partial p_{j}} \left( \frac{\partial^{2} S}{\partial t \partial q_{j}} + \frac{\partial^{2} S}{\partial q_{k} \partial q_{j}} \frac{\partial H}{\partial p_{k}} + \frac{\partial H}{\partial q_{j}} \right), \qquad i = 1, \ldots, n.
\end{eqnarray*}
By virtue of (\ref{reg}), the last equations are equivalent to
\begin{eqnarray*}
0 & = & \frac{\partial^{2} S}{\partial t \partial q_{j}} + \frac{\partial^{2} S}{\partial q_{k} \partial q_{j}} \frac{\partial H}{\partial p_{k}} + \frac{\partial H}{\partial q_{j}} \\
& = & \frac{\partial}{\partial q_{j}} \left[ \frac{\partial S}{\partial t} + H(q_{k}, \frac{\partial S}{\partial q_{k}}, t) \right], \qquad j = 1, \ldots, n,
\end{eqnarray*}
which implies that the expression inside the brackets is a function of $t$ only,
\[
\frac{\partial S}{\partial t} + H(q_{k}, \frac{\partial S}{\partial q_{k}}, t) = \chi(t).
\]
Thus,
\[
\tilde{S} = S - \int^{t} \chi(u) \, {\rm d} u
\]
is a solution of the HJ equation. We can verify that this solution is complete by differentiating (\ref{eigf}) with respect to $\lambda_{j}$, which gives
\[
\frac{\partial F_{i}}{\partial p_{k}} \frac{\partial^{2} S}{\partial \lambda_{j} \partial q_{k}} = \delta_{ij}.
\]
Taking into account (\ref{reg}), this last equation shows that $\det (\partial^{2} S/\partial \lambda_{j} \partial q_{k}) \not= 0$.

\section{Examples}
In this section we give some examples of the method presented above.

\subsection{One-dimensional harmonic oscillator}
The function
\[
F(q, p, t) = m \omega q \sin \omega t + p \cos \omega t
\]
already considered above [see (\ref{cons})], is a constant of motion if the Hamiltonian is given by
\begin{equation}
H = \frac{p^{2}}{2m} + \frac{m \omega^{2}}{2} q^{2}, \label{hoh}
\end{equation}
where $\omega$ is a constant. According to the results of the preceding section, if $\lambda$ is a constant,
\begin{equation}
S = \lambda q \sec \omega t - \frac{m \omega}{2} q^{2} \tan \omega t + \phi(t)
\end{equation}
must be a solution of the HJ equation for the Hamiltonian (\ref{hoh}), if the function $\phi$ is appropriately chosen. A direct computation yields
\[
\frac{1}{2m} \left( \frac{\partial S}{\partial q} \right)^{2} + \frac{m \omega^{2}}{2} q^{2} + \frac{\partial S}{\partial t} = \frac{\lambda^{2}}{2m} \sec^{2} \omega t + \phi'(t),
\]
and, therefore, choosing $\phi(t) = - \lambda^{2} \tan \omega t/2m \omega$, we obtain the complete solution of the HJ equation
\[
S  = \lambda q \sec \omega t - \left( \frac{\lambda^{2}}{2m} + \frac{m \omega^{2}}{2} q^{2} \right) \frac{\tan \omega t}{\omega}.
\]

Note that, in this case, $H$ is also a constant of motion and, as pointed out above, the equation that determines the eigenfunctions of $H$ is just the time-independent HJ equation. However, the constant of motion (\ref{cons}) leads to simpler expressions.

\subsection{Particle in a time-dependent force field}
As a second example we consider the time-dependent Hamiltonian
\[
H = \frac{p^{2}}{2m} - ktq,
\]
where $k$ is a constant. One can readily verify that
\[
F = p - \frac{kt^{2}}{2}
\]
is a constant of motion and that the eigenfunctions of $F$, i.e., the solutions of
\[
\frac{\partial S}{\partial q} - \frac{kt^{2}}{2} = \lambda,
\]
are given by
\begin{equation}
S = \lambda q + \frac{kt^{2}}{2} q + \phi(t), \label{pf2}
\end{equation}
where $\phi(t)$ is an arbitrary function of $t$ only. Then
\[
\frac{1}{2m} \left( \frac{\partial S}{\partial q} \right)^{2} - \frac{kt^{2}}{2} + \frac{\partial S}{\partial t} = \frac{\lambda^{2}}{2m} + \frac{\lambda k t^{2}}{2m} + \frac{k^{2} t^{4}}{8m} + \phi'(t),
\]
hence, choosing
\[
\phi(t) = - \frac{\lambda^{2} t}{2m} - \frac{\lambda kt^{3}}{6m} - \frac{k^{2} t^{5}}{40m},
\]
(\ref{pf2}) is a complete solution of the HJ equation (which is not separable).

\subsection{Particle in two dimensions}
As a final example we consider the Hamiltonian
\begin{equation}
H = \frac{1}{2m} \left[ \left( p_{x} + \frac{eB}{2c} y \right)^{2} + \left( p_{y} - \frac{eB}{2c} x \right)^{2} \right], \label{magn}
\end{equation}
which corresponds to a charged particle of mass $m$ and electric charge $e$ in a uniform magnetic field $B$. The functions
\begin{eqnarray}
F_{1} & = & \frac{1}{2} (1 + \cos \omega t) p_{x} - \frac{1}{2} p_{y} \sin \omega t + \frac{m \omega}{4} x \sin \omega t - \frac{m \omega}{4} (1 - \cos \omega t) y, \label{f1} \\
F_{2} & = & \frac{1}{2} (1 + \cos \omega t) p_{y} + \frac{1}{2} p_{x} \sin \omega t + \frac{m \omega}{4} y \sin \omega t + \frac{m \omega}{4} (1 - \cos \omega t) x, \label{f2}
\end{eqnarray}
where $\omega \equiv eB/mc$, are constants of motion in involution, which correspond to the values of the canonical momenta $p_{x}$ and $p_{y}$, respectively, at $t = 0$.

From (\ref{f1}) and (\ref{f2}) one finds that the common eigenfunctions of $F_{1}$ and $F_{2}$, with eigenvalues $\lambda_{1}$ and $\lambda_{2}$, respectively, are
\[
S = \lambda_{1} x + \lambda_{2} y + \tan {\textstyle \frac{1}{2}} \omega t \left[ \lambda_{2} x - \lambda_{1} y - \frac{m \omega}{4} (x^{2} + y^{2}) \right] + \phi(t),
\]
where $\phi(t)$ is an arbitrary function of $t$ only. Substituting this expression into the HJ equation one finds that $S$ is a solution of this equation if and only if
\[
\frac{\lambda_{1}{}^{2} + \lambda_{2}{}^{2}}{2m} \sec^{2} {\textstyle \frac{1}{2}} \omega t + \phi'(t) = 0,
\]
hence,
\[
S = \lambda_{1} x + \lambda_{2} y - \left[ \left( \lambda_{1} + \frac{m \omega}{2} y \right)^{2} + \left( \lambda_{2} - \frac{m \omega}{2} x \right)^{2} \right] \frac{\tan {\textstyle \frac{1}{2}} \omega t}{m \omega}
\]
is a complete solution of the HJ equation.

\section{Concluding remarks}
As pointed out above, the formulation of the Liouville theorem given here makes use of terms analogous to those employed in the standard formalism of quantum mechanics, thus providing another example of the parallelism between both theories. Another advantage of the version of the Liouville theorem given above is that its proof is shorter than those usually presented in the textbooks.

\end{document}